\documentstyle{article}
\font\tenrm=cmr10
\font\tenit=cmti10
\font\elevenbf=cmbx10 scaled\magstep 1
\font\elevenrm=cmr10 scaled\magstep 1

\renewenvironment{thebibliography}[1]
 { \elevenrm
   \begin{list}{\arabic{enumi}.}
    {\usecounter{enumi}     \setlength{\parsep}{0pt}
     \setlength{\itemsep}{3pt} \settowidth{\labelwidth}{#1.}
     \sloppy
    }}{\end{list}}

\begin{document}
\begin{center}{\elevenbf MULTISKYRMIONS AND BARYONIC BAGS.}\\
\vglue 0.5cm
{\tenrm Vladimir B.Kopeliovich \\}
{\tenit Institute for Nuclear Research of the Russian Academy of
Sciences, Moscow 117312, Russia\\}
\end{center}
\vglue 0.3cm
{\rightskip=2pc
 \leftskip=2pc
\tenrm\baselineskip=11pt
 \noindent
Analytical treatment of skyrmions given by rational map ($RM$) ansaetze
proposed recently for the Skyrme model is extended for the model including
the $6$-th order term in chiral field derivatives in the lagrangian 
and used for the calculations of different properties of
multiskyrmions. At large baryon numbers the approximate
solutions obtained are similar to the domain wall, or
to spherical bubbles with energy and baryon number density concentrated at 
their boundary. Rigorous upper bound is obtained for the masses of $RM$ 
multiskyrmions which is close to the known masses, especially at large 
$B$. For the $6$-th order variant the lower bound for masses of $RM$
skyrmions is
obtained as well. The main properties of the bubbles of matter are obtained for 
arbitrary number of flavours. They are qualitatively the same for the $4$-th 
and $6$-th order terms present in the lagrangian, although differ 
in some details.
\vglue 0.6cm}
\elevenrm
\baselineskip=15pt
\section{Introduction} 
Soliton models of different kinds are used in various fields of
physics. In elementary particle physics the soliton models provide a concept
of baryons as extended in space objects, opposite to the concept
of the point-like objects, usual for the quantum field theory.
The chiral soliton approach, starting with several basic 
principles and ingredients incorporated in the model lagrangian \cite{1,2}
provides realistic and even satisfactory description of baryons and baryonic
systems. The latter are obtained within this approach
as quantized solitonic solutions of equations of motion, characterized by the
so called winding number or topological charge which is identified with
the baryon number $B$.
Numerical studies have shown that the chiral field configurations of lowest
energy possess different topological 
properties - the shape of the mass and $B$-number distribution - for different 
values of $B$. It is a sphere for $B=1$ hedgehog \cite{1}, a torus for $B=2$,
tetrahedron for $B=3$, cube for $B=4$, and higher
polyhedrons for greater baryon numbers. The symmetries of
various configurations for $B$ up to $22$ and their masses have been 
determined in \cite{3} (the references to earlier original papers where the
symmetries of configurations with smaller baryon numbers have been determined 
can be found in \cite{3,4}). These configurations have one-shell structure and 
for $B > 6$ all of them, except two cases, are
formed from $12$ pentagons and $2B-14$ hexagons; in carbon chemistry
similar structures are known as fullerenes \cite{3}. The mass and baryon number
densities for these configurations are concentrated along the edges of
polyhedrons. All these 
configurations can be made of $2B-2$ slightly deformed torus-like 
configurations glued together, which can be considered by this reason as 
elementary building blocks for multiskyrmions. As will be shown here, the 
dimensions of these elementary cells do not depend on $B$ when $B$ is large 
enough.

The so called rational map $(RM)$ ansatz, proposed for the $SU(2)$ skyrmions in 
\cite{5} and widely used now, in present paper as well, allows to simplify the 
problem of finding the configurations
of lowest energy. For the $RM$ ansatz the minimization of the skyrmions energy
functional proceeds in two steps: at first step the map from $S^2 \to S^2$ is
minimized for the $SU(2)$ model (for the $SU(N)$ model it is a map from $S^2 \to 
CP^{N-1}$, \cite{6}), and, second, the energy functional depending on skyrmion 
profile as a function of distance from center of skyrmion is minimized.
As will be shown here, just the second step can be done analytically with
quite good accuracy. Many important properties of the $RM$ multiskyrmions can
be studied in this way, and some of them do not
depend on result of the first step. This allows to make certain conclusions 
for the arbitrary large $B$ and for any number of flavours $N_F=N$ independently of 
presence of numerical calculations. Without difficulties the consideration
has been extended to the variant of the model with the $6$-th order terms
in chiral derivatives included into lagrangian (the SK6 variant of the Skyrme
model). Remarkably, that for the SK6 variant of the model the dependence of
the results on the first step of calculation is even weaker than for the SK4
variant.

Beginning with \cite{1,2}, the chiral soliton models have been considered as a 
special class of models for baryons. Their connection with other models
could be instructive and useful, and this is also an issue of present 
paper. In particular, it is shown that Skyrme-type models provide field
theoretical realization of the bag model of special kind for baryonic systems.\\

\section{Large $B$ multiskyrmions as spherical bubbles or domain walls}
Here we consider the multiskyrmions in the general $SU(N)$ case; detailed
comparison of analytical results
with numerical calculations is made in the $SU(2)$ model and also in the $SU(3)$ 
variant using the projector ansatz \cite{6}.
In the $SU(2)$ model the chiral fields are functions of the profile $f$ and the 
unit vector
$\vec{n}$, according to definition of the unitary matrix $U \in SU(2)$
$U=c_f+is_f\vec{n}\vec{\tau}$.
For the ansatz based on the rational maps the profile
$f$ depends only on the variable $r$, and the components of vector $\vec{n}$ - on
angular variables $\theta, \, \phi$. 
$n_1=(2\, Re\, R)/(1+|R|^2) ,\; n_2=(2\, Im\, R)/(1+|R|^2) ,\; n_3 =(1-|R|^2)/
(1+|R|^2)$, where $ R$ is a rational function of variable $z=tg(\theta/2)
exp(i\phi)$ defining the map of degree ${\cal N}$ from $S^2 \to S^2$.

The notations are used \cite{5}
$$ {\cal N} = {1\over 8\pi}\int r^2(\partial_i \vec{n})^2 d\Omega =
{1\over 4\pi}\int \frac{2i dR d\bar{R}}{(1+|R|^2)^2} $$
$${\cal I} = {1 \over 4\pi}\int r^4\frac{[\vec{\partial}n_1\vec{\partial}n_2]^2}
{n_3^2} d\Omega = {1 \over 4\pi} \int \Biggl(\frac{(1+|z|^2)}{(1+|R|^2)}
{|dR| \over |dz|}\Biggr)^4 \frac{2i dz d\bar{z}}{(1+|z|^2)^2}, \eqno (1) $$
where $\Omega$ is a spherical angle.
For $B=1$ hedgehog ${\cal N}={\cal I} =1$. ${\cal N}=B$ for configurations of
lowest energy.

For more general $SU(N)$ case and using projector ansatz one obtains \cite{6}
$${\cal N} = {i \over 2\pi}\int dz d\bar{z} Tr[\partial_zP\partial_{\bar{z}}P],
$$
$${\cal I} = {i \over 4\pi}\int dz d\bar{z} (1+|z|^2)^2Tr[\partial_zP,
\partial_{\bar{z}}P]^2, \eqno (2)$$
$P$ is a projector, $N$x$N$ hermitian matrix, $P=f\,f^\dagger/f^\dagger f$,
$Tr P=1$. For $SU(2)$ case the $2$-component column
$f = (R,\, 1)^T$, and $(1)$ can be obtained easily from formulas $(2)$.

The classical mass of skyrmion for $RM$ ansatz in universal 
units $3\pi^2 F_\pi/e$ is \cite{6,7}:
$$ M={1 \over 3\pi}\int \biggl\{A_N r^2f'^2 +2Bs_f^2[1+(1-\lambda)f'^2] + 
(1-\lambda) {\cal I} \frac{s_f^4}{r^2} +\lambda {\cal I} \frac{s_f^4}{r^2}f'^2 
\biggr\} dr,  \eqno (3) $$
$r$ measured in units $2/(F_\pi e)$,
the coefficient $A_N = 2(N-1)/N$ for symmetry group $SU(N)$ \cite{6}, and this 
generalization provides a possibility to consider models with arbitrary number of
flavours $N=N_F$ - essentially nonembeddings of $SU(2)$ in $SU(N)$.
$\lambda$ defines the weight of the $6$-th order term. If $\lambda =0$ we
obtain the original Skyrme model variant (we call it the SK4-variant), $\lambda =1$
corresponds to the pure SK6-variant.
The expression $(3)$ for mixed variant was obtained in \cite{7},
where the properties of multiskyrmions with $B \leq 5$ have been 
investigated in the modification of the Skyrme model with $6-th$ order
term included into lagrangian.
For $B=2$ the results obtained in \cite{7} agree with those obtained in \cite{8}
which correspond to the value of $\lambda =1$. Besides axially symmetrical
configurations with $B\leq 5$ in the SK4 variant the torus-like configuration with
$B=2$ was obtained in \cite{8} also in the SK6 variant. For reasonable values
of model parameters the binding energy was about twice greater than that
for pure Skyrme model: it was found
$\sim 150\, Mev$ instead of $74\, Mev$ for the variant of the SK4 model which
provides fit for the masses of nucleon and $\Delta$-isobar.

It was important in deriving $(3)$
that the integral over angular variables for the trace of the third power of 
$[\partial_zP,\partial_{\bar{z}}P]$ equals to zero \cite{7}:
$$ {\cal M} = {i\over 8\pi}\int dz d\bar{z} (1+|z|^2)^4 Tr[\partial_zP
\partial_{\bar{z}}P]^3 = 0, $$
therefore, the structure $(3)$ is specific for the rational map ansatz.

To find the minimal energy configuration at fixed
${\cal N}=B$ one minimizes ${\cal I}$, and then finds the profile $f(r)$ by 
minimizing energy $(3)$. 
The inequality takes place: ${\cal I} \geq B^2$ \cite{5,6}.
The important consequence of $(3)$ is that the symmetries of multiskyrmions 
in the SK6 variant \cite{7} are the same as for the SK4 variant, because the quantity 
${\cal I}$ is the same for both variants.

Numerical calculations peformed in \cite{3}-\cite{6} have shown, and the 
analytical treatment of \cite{9} and here supports, that at large $B$ and, 
hence, large ${\cal I}$ the multiskyrmion looks
like a spherical bubble with profile equal to $f=\pi$ inside and $f=0$
outside. The energy and $B$-number density of this configuration is concentrated 
at its shell, similar to the domain walls system considered in \cite{10}
in connection with cosmological problems.

The lower bound for the mass of solitons can be obtained from $(3)$ for the SK6
variant, similar to the SK4 variant, known previously.
Using evident relations $A_Nf'^2r^2 + {\cal I}s_f^4 f'^2/r^2 \geq 2\sqrt{A_N
{\cal I}} f'^2 s_f^2$, and $f'^2\sqrt{A_N{\cal I}}+B \geq -2f'\bigl(B\sqrt{A_N
{\cal I}}\bigr)^{1/2}$ (recall that $f' <0$) we obtain from $(3)$ for arbitrary
value of real positive parameter $\lambda \leq 1$:
$${M(B,\lambda) \over B} \geq {2\lambda\over 3}
\biggl(\frac{A_N{\cal I}}{B^2}\biggr)^{1/4}
\,+\, {1-\lambda\over 3}\biggl[2+\biggl(\frac{A_N{\cal I}}{B^2}\biggr)^{1/2}
\biggr]. \eqno (4a) $$
For $\lambda =0$ the known results of \cite{5,6} are reproduced, for the SK6
variant ($\lambda =1$) it is new bound.
When the last term in $(3)$ can be neglected, another lower bound can be
obtained, see also \cite{7}:
$${M(B,\lambda) \over B} \geq  {\sqrt{1-\lambda}\over 3}
\biggl[2+\biggl(\frac{A_N{\cal I}}{B^2}\biggr)^{1/2} \biggr], \eqno (4b) $$
which does not provide real restriction when $\lambda $ is close to $1$.
Recall that the bound ${M / B} > 1$ was obtained for arbitrary, not
only $RM$, $SU(2)$ skyrmions first by Skyrme in the SK4 variant of the model 
\cite{1}.

Denote $\phi=cos f$, then the energy $(3)$ can be presented as
$$M={1\over 3\pi}\int \biggl\{{1 \over (1-\phi^2)}
\bigl[A_Nr^2\phi'^2+2B(1-\phi^2)^2\bigr] +(1-\lambda)\bigl[2B\phi'^2+
{\cal I}(1-\phi^2)^2/r^2\bigr]+$$
$$+\lambda {\cal I}(1-\phi^2)\phi'^2/r^2 \biggr\}dr,  \eqno (5) $$
with $\phi$ changing from $-1$ at $r=0$ to $1$ at $r\to\infty$. 
The first part of $(5)$ is the second order term contribution into the mass,
the second - the Skyrme term contribution, and the last, proportional to
$\lambda$, - the 6-th order term. At fixed $r=r_0$ the $4$-th order term 
is proportional exactly to $1$-dimensional domain wall energy. 
It is possible to write the second order contribution in $(5)$ in the form:
$$M^{(2)}={1\over 3\pi}\int \biggl\{{A_Nr^2 \over (1-\phi^2)}
\biggl[\phi'- \sqrt{{2B\over A_N}}(1-\phi^2)/r\biggr]^2 + 2r\sqrt{2A_NB}\phi' 
\biggr\}dr, $$
and similar for the $4$-th order Skyrme term. The equality $\phi'=\sqrt{2B/A_N}
(1-\phi^2)/r$ eliminates considerable part of integrand for $M^{(2)}$.
Therefore, it is natural to consider function $\phi$ satisfying the following
differential equation:
$$ \phi'= {b \over 2r}(1-\phi^2) \eqno (6) $$
with constant power $b$, which has solution satisfying boundary conditions 
$\phi(0)=-1$ and $\phi(\infty )=1$:
$$ \phi (r,r_0,b) = \frac{(r/r_0)^b-1}{(r/r_0)^b+1} \eqno (7)$$
with arbitrary $r_0$ - the distance from the origin of the point where 
$\phi=0$ and profile $f=\pi/2$. $r_0$ can be considered as a radius of
multiskyrmion, both $b$ and $r_0$ will be defined further by means of the
mass minimization procedure. The radii of distributions of baryon number and 
mass in the soliton are close to $r_0$, see Section 5 below.

After substitution of this ansatz one obtains the soliton mass
in the form:
$$M(B,b) = {1 \over 3\pi}\int \biggl\{(A_Nb^2/4+2B)(1-\phi^2)
+ (1-\lambda)(Bb^2/2+{\cal I})(1-\phi^2)^2/r^2 +$$
$$+\lambda {\cal I} b^2(1-\phi^2)^3/(4r^4) \biggr\} dr \eqno (8) $$
Integrating over $dr$ can be made using known
expressions for the Euler-type integrals, e.g. 
$$\int_0^\infty \frac{dr}{1+(r/r_0)^b}=\frac{\pi r_0}{b\, sin(\pi/b)},
\qquad b>1, $$ 
and, more generally
$$\int_0^\infty \frac{(r/r_0)^cdr}{\beta+(r/r_0)^b}=\beta^{(1+c-b)/b}
\frac{\pi r_0}{b\, sin[\pi(1+c)/b]},
\qquad \beta > 0, \;b > 1+c,\; c > -1 .\eqno (9) $$
Differentiation in $\beta$ allows to get the integral with any power of
$1+(r/r_0)^b$ in denominator.
What we need now is ($\phi $ is defined in $(7)$)
$$ \int (1-\phi^2)\, dr = \frac{4\pi r_0}{b^2 sin(\pi/b)}, \qquad
\int \frac{(1-\phi^2)^2}{r^2} \,dr \;=\;\frac{8\pi(1-1/b^2)}{3r_0b^2sin(\pi/b)},$$
$$ \int \frac{(1-\phi^2)^3}{r^4} \,dr \;=\;\frac{32\pi(1-9/b^2)(1-9/4b^2)}
{5r_0^3b^2sin(3\pi/b)}= \frac{32\pi}{15r_0^3b^2sin(\pi/b)}
 F_3(b), \eqno (10) $$
with $F_3(b)=3 \,sin(\pi/b)(1-9/b^2)(1-9/4b^2)/sin(3\pi/b)$.
Asymptotically at large $b$ the function $F_3(b) \to 1$, it is really close to 
$1$ in wide interval of argument, from $b\sim 6$ up to $\infty$, the deviation does not
exceed several $\%$. However, correction due to deviation of $F_3$ from $1$
will be taken into account.

Expressions $(9,10)$ allow to obtain the mass of multiskyrmion in simple 
analytical form as a function of parameters $b$ and $r_0$:
$$M(B,r_0,b)= \frac{1}{3b\,sin(\pi/b)}\biggl[ (A_Nb^2+8B)
{r_0\over b} +{4\over 3br_0}(1-\lambda)(Bb^2+2{\cal I})(1-1/b^2)
+\lambda {\cal I} {8b\over 15r_0^3} F_3(b)\biggr]. \eqno (11)$$
Since $b \sim 2\sqrt{B}$, or greater, see below, we can put $F_3 \to 1$ and 
substitute at large enough $B$ $\pi/b\,sin(\pi/b) \to 1$, to obtain:
$$M(B,r_0,b)\simeq {1\over 3\pi}\biggl\{ b\biggl[A_Nr_0+(1-\lambda)
{4B\over 3r_0}+\lambda \frac{8{\cal I}}{15r_0^3}\biggr] +{8\over b}
\biggl[Br_0\,+(1-\lambda )\frac{{\cal I}}{3r_0} \biggr] \biggr\}. \eqno (12) $$
Now minimization over $b$ can be done without
difficulties. It provides an upper bound for the soliton mass, because we
restricted ourselves with the profiles of the type $(7)$, only. 
Minimizing in $b$ is trivial, giving
$$b_{min}= \sqrt{G_N/G_D},\qquad M(B,r_0)={2\over 3\pi}\sqrt{G_NG_D}
\eqno (13)$$
with
$$G_N = 8\biggl[Br_0+(1-\lambda)\frac{{\cal I}}{3r_0}\biggr],\qquad
G_D=A_Nr_0+(1-\lambda){4B\over 3r_0}+\lambda {\cal I}{8\over 15r_0^3}.
\eqno(14)$$
Minimization of the mass in $r_0$ provides now $r_0^{min}$ and upper bound
for the mass $M(B)$.
The extreme cases $\lambda =0$ (SK4) and $\lambda =1$ (SK6) are simple, and
we present here results for both cases.\\
The SK4 case, considered in \cite{9}.
$$r_0^{min}\simeq \biggl({2\over 3}\sqrt{\frac{\cal I}{A_N}}\biggr)^{1/2}, 
\; b^{min}_0 \simeq 2({\cal I}/A_N)^{1/4},\qquad
{M\over B} < {4\over 3\pi}\biggl({2\over 3}\biggr)^{1/2}
(2+\sqrt{A_N{\cal I}/B^2}) \eqno (15) $$
The SK6 case.\\
$$r_0^{min} \simeq \biggl(\frac{8{\cal I}}{15\,A_N}\biggr)^{1/4}, \qquad
b^{min}_0 =2\sqrt{B/A_N},\qquad
{M\over B} < {8 \over 3\pi}\biggl(\frac{8A_N{\cal I}}{15B^2}
\biggr)^{1/4}. \eqno (16) $$
The masses of multiskyrmions in the SK6 case at large $B$ are smaller
than those in the SK4 case, the ratio of radii is greater: $r_0(SK6)/r_0(SK4) =
(6/5)^{1/4} \simeq 1.0466$.

For the SK4 variant it is possible to obtain more accurate estimates describing
also preasymptotics in $B$.
$$r_0^{min}=2 \biggl[\frac{(Bb^2+2{\cal I})(1-1/b^2)}{3(A_Nb^2+8B)}
\biggr]^{1/2} \eqno (17) $$
and
$$M(B,b)/B= \frac{4}{3b\,sin(\pi/b)}\bigl[(b^2+2{\cal I}/B)(A_N b^2+8B)
(1-1/b^2)/(3b^2B)\bigr]^{1/2} \eqno (18) $$

At large enough $B$ when it is possible to neglect the influence of slowly
varying factors $(1-1/b^2)$ and $b\,sin(\pi/b)$ we obtained \cite{9}
$$b^{min}=b_0= 2({\cal I}/A_N)^{1/4},\qquad
r_0^{min}\simeq \biggl[{2\over 3}\biggl(\sqrt{\frac{{\cal I}}{A_N}}-
{1\over 4}\biggr)\biggr]^{1/2} \eqno (19)$$
and, therefore, approximately
$${1\over 3}(2+\sqrt{{\cal I}A_N}/B) < {M\over B} < {1 \over 3}(2 + 
\sqrt{{\cal I}A_N}/B)
\frac{4}{b_0 sin(\pi/b_0)}\biggl[{2\over 3}\biggl(1-
{1\over b_0^2}\biggr)\biggr]^{1/2}. \eqno (20) $$
The lower bound in $(20)$ is taken from $(4a)$.

The correction to the value $b_0$ can be found including into minimization
procedure the factor $(1-1/b^2)$ and variation of $b\, sin (\pi/b)\simeq 
\pi [1-\pi^2/(6b^2)] $.
It provides:
$$\delta \, b \simeq \frac{B(\pi^2/3-1)(2+\sqrt{A_N{\cal I}}/B)^2}
{16 {\cal I}^{3/4}A_N^{1/4}}, \eqno (21) $$
and the value $b=b_0+\delta\,b $ should be inserted into $(18)$. This improves
the values of $M/B$ for $B=1,2,3... $ but provides negligible effect for 
$b \sim 17$ and greater, since $\delta\,b \sim 1/\sqrt{B}$.
The comparison of numerical calculation result and analytical upper bound
$(18)$ is presented in {\bf Table 1}.

For the SK6 variant from $(11)$ we have after minimization in $r_0$
$$ (r_0^{min})^2=\biggl[\frac{8{\cal I}F_3(b)b^2}{5(A_Nb^2+8B)}\biggr]^{1/2}
\eqno (22) $$
and
$${M(B,b)\over B}= \frac{4}{9b\,sin(\pi/b)}\biggl[\frac{8{\cal I}F_3(b)}
{5Bb^2}(A_N b^2/B+8)^3\biggr]^{1/4} \eqno (23) $$
which provides the upper bound for the soliton mass, for any $b$, similar to
$(18)$.
Combining with $(4a)$, we obtain approximately
$${2\over 3}\biggl(\frac{{\cal I}A_N}{B^2}\biggr)^{1/4} < {M\over B} < 
{8 \over 3b_0\,sin(\pi/b_0)}
\biggl(\frac{8A_N{\cal I}F_3(b)}{15B^2}\biggr)^{1/4}. \eqno (24) $$
The correction to the value of $b_0=2\sqrt{B/A_N}$ can be taken into account 
similar to the SK4 case,
$$\delta \, b(SK6) \simeq \frac{3(2\pi^2-45/4)}{4\, b}. \eqno (25) $$
The results for the upper bound for the masses of the SK6 multiskyrmions calculated
according to $(23)$ are given in {\bf Table 2}, next Section.

A comment concerning the behaviour of the profile $f$ at large $r$ is 
necessary. It is well known that asymptotically at $r\to \infty$ $f$ is defined
by the $2-d$ order term in the lagrangian and is proportional to $f \sim 1/r^p$
with $p=1/2 + \sqrt{2B+1/4} $. So, $p=2$ for $B=1$, $p=3$ for $B=3$, etc.
Obviously, the tail of the profile we obtained here, $f \sim 1/r^{b/2}$ with
$b$ given in $(15)$ or $(16)$, is greater and falls down more slowly than the
true one. This is because our purpose is to describe the masses of skyrmions
and other global characteristics, but not the asymptotic behaviour of the
profile. \\
\section{Numerical results and comparison of the SK6 and SK4 variants of the 
          model}
Numerically $(18)$ provides the upper bound for the skyrmion masses which differs 
from the masses of known $RM$ multiskyrmions in the SK4 variant within $\sim 2$\%, 
beginning with $B=2$, see {\bf Table 1}. The values $M/B|_{RM}$ are calculated
numerically by means of minimization of the functional $(3)$.
Even for $B=1$, where the method evidently should not work well, we obtained
$M=1.271$ for $b=2.85$ instead of precise value $M=1.232$. It should be noted
here that analytical results of paper \cite{9}
for smallest $B$ are slightly improved here due to 
better choice of power $b(B)$: it is found by numerical minimization of $(18)$,
whereas in \cite{9} we used approximate formula $(21)$ for $\delta b$.
For maximal values of $B$ between $17$ and $22$ where the value of ${\cal I}$ 
is calculated, the upper bound exceeds the $RM$ value of mass by $0.5 \%$ only.
We took here the ratio $R_{I/B}={\cal I}/B^2$ in the cases where
this ratio is not determined yet, the same as for highest $B$ where it is known,
i.e. $1.28$ for $SU(2)$ case \cite{3}, $B=32$ and $64$, and $1.037$ for $B>6$ 
in $SU(3)$ \cite{6}. 
\begin{center}
\begin{tabular}{|l|l|l|l|l|l|l|l|l|l|l|l|}
\hline
 $B$ &$2 $ &$3 $&$4  $&$5 $ &$6 $&$7$&$13$&$17$&$22$&$32$&$64$ \\
\hline
$M/B|_{RM}$  &$1.208$&$1.184$&$1.137$&$1.147$&$1.137$&$1.107$&
$1.099$&$1.091$&$1.092$&$1.088$&$1.084$ \\
\hline
$r_B$&$1.45$&$1.73$&$1.89$&$2.13 $&$2.30$&$2.40 $&$3.23$&$3.65 $&
$4.15$&$4.97 $&$6.98 $\\
\hline
$b(B)$&$3.73$&$4.38$&$4.77$&$5.35$&$5.76$&$6.00$&$7.98$&$9.01$&$10.23$&$12.24$&
$17.16$\\
\hline
$M/B|_{appr}$&$1.227$&$1.198$&$1.150$&$1.158$&$1.147$ &$1.117$&$1.106$&
$1.0976$&$1.098$&$1.094$&$1.089$ \\
\hline
$M/B|_{num}$&$1.1791$&$1.1462$&$1.1201$&$1.1172$&$1.1079$&$1.0947$&
$1.0834$&$1.0774$&$1.0766$&---&--- \\
\hline
\hline
$M/B|^{SU_3}_{RM}$ &$1.222$&$1.215$&$1.184$&$1.164$&$1.145$&$1.138$&$1.120$&
$1.115$&$1.111$&$1.1064$&$1.101$ \\
\hline
$r_B^{SU3}$&$1.28$&$1.54$&$1.72$&$1.88$&$2.02 $&$2.16 $&$2.87$&$3.25$&
$3.68$&$4.40$&$6.18 $\\
\hline
$b(B)^{SU_3}$&$3.32$&$3.96$&$4.38$&$4.76$&$5.08$&$5.43$&$7.12$&$8.05$&$9.08$&
$10.86$&$15.19$
\\
\hline
$M/B|^{SU_3}_{appr}$&$1.247$&$1.231$&$1.198$&$1.176$&$1.156$&$1.149$&$1.127$&
$1.121$&$1.116$&$1.111$&$1.106$ \\
\hline
\end{tabular}
\end{center}
{\tenrm
\baselineskip=11pt {\bf Table 1.} The skyrmion mass per unit $B$-number
in universal units $3\pi^2F_\pi/e$
for the $RM$ configurations, the SK4 variant, approximate and precise solutions. 
$r_B = \sqrt{<r^2_B>}$ - mean square radius of baryon number distribution,
or the isoscalar radius, in units $2/F_\pi e$. The approximate values 
(upper bounds) are calculated using
formula $(18)$ with the power $b$ minimizing it. The numerical values for the
$SU(2)$ model are from
the papers \cite{3} and earlier papers. The last $4$ lines show the result for 
the $SU(3)$ projector ansatz \cite{6} and approximation to this case, $A_N=4/3$.
Calculations of $M/B|_{appr}^{SU3}$ are made also with the help of $(18)$ with 
the power $b(B)^{SU3}$ which minimizes it.}\\

It is of interest to compare the same quantities $(b_0, \;r_0,\; M/B)$ at large
baryon numbers for the SK4 variant (original Skyrme model) and the SK6 variant.
$$b_0^2(SK6)/b_0^2(SK4)=[B^2/({\cal I}A_N)]^{1/2}\,<\,1, \eqno (26)$$
$$ r_0^2(SK6)/r_0^2(SK4) =(6/5)^{1/2} \, >\,1. \eqno (27) $$
          
Asymptotically at large $B$ the ratio of upper and lower bounds \cite{9}
$$ R_{max/min}(SK4)={M_{max}\over M_{min}}|_{SK4} = {4\over \pi} 
\biggl({2\over 3}\biggr)^{1/2} \simeq 1.0396,\eqno (28) $$
i.e. the gap between upper and lower bounds is less than $4 \%$,
independently on $B$, the particular value of ${\cal I}$ and the number of 
flavours $N$.
For the SK6 variant we obtain
$$ R_{max/min}(SK6)={M_{max}\over M_{min}}|_{SK6} = {4\over \pi} 
\biggl({8\over 15}\biggr)^{1/4}\simeq 1.0881,\eqno (29) $$
here the gap is less than $9\%$, also independently on $B,\,{\cal I}$ and $N$.

Note that the ratio
$$ R_{max/min}(SK6)/R_{max/min}(SK4)=
r_0(SK6)/r_0(SK4)\simeq (6/5)^{1/4} \simeq 1.0466, \eqno (30) $$
so, the number $(6/5)^{1/4}$ plays a special role in comparison of both 
variants of the model.

The SK6 variant reveals even weaker dependence on the quantity ${\cal I} $ 
than the SK4 variant, the power $b_0$ does not depend on ${\cal I}$ at all, cf.
$(15)$ and $(16)$ above.

With decreasing ${\cal I}$ the upper bounds decrease proportionally to
the lower bounds. It should be stressed that our upper bounds for masses of
the SK4 and SK6 multiskyrmions are obtained on a class of profile functions $(7)$,
and, probably, can be improved.
\begin{center}
\begin{tabular}{|l|l|l|l|l|l|l|l|l|l|l|l|}
\hline
 $B$ &$2 $ &$3 $&$4  $&$5 $ &$7$&$13$&$17$&$22$&$32$&$64$&$128$ \\
\hline
$M/B|^{SK6}_{RM}$&$0.9181$&$0.8843$&$0.8289$&$0.837$&$0.793$&$0.781$&$0.772$&
$0.773$&$0.768$&$0.763$&$0.760$ \\
\hline
 $b(B)$ &$4.08$&$4.66$&$5.14$&$5.55$ &$6.27$&$8.00$&$8.95$&$10.01$&$11.85$&
$16.39$&$22.9$ \\
\hline
$M/B|^{SK6}_{appr}$&$0.951$&$0.915$&$0.857$&$0.865$&$0.817$&$0.801$&$0.791$&
$0.790$&$0.784$&$0.778$&$0.774$ \\
\hline
$r_{B,RM}^{SK6}$&$1.636$&$1.910$&$2.047$&$2.293$&$2.540$&$3.351$&$3.767$&
$4.269$&$5.093$&$7.125$&$10.037$ \\
\hline
\hline
$M^{SK6}_{RM}/M^{SK4}_{RM}$&$0.760$&$0.747$&$0.729$&$0.730$&$0.716$&$0.711$&
$0.708$&$0.7075$&$0.706$&$0.7044$&$0.704$ \\
\hline
$r_B^{SK6}/r_B^{SK4}$&$1.128$&$1.107$&$1.082$&$1.076$&$1.056$&$1.039$&
$1.033$&$1.029$&$1.025$&$1.022$&$1.019$ \\
\hline
\end{tabular}
\end{center}
{\tenrm
\baselineskip=11pt {\bf Table 2.} 
The skyrmion mass per unit $B$-number for the SK6 variant of the model
($\lambda = 1$) with $SU(2)$ flavour symmetry, in units $3\pi^2F_\pi/e$, 
and comparison with the SK4 variant. The approximate upper bound $M/B|_{appr}$ is 
calculated using formulas $(23)$ with the power $b(B)$ given in third
line. The isoscalar radius $r_{B,RM}$ in units $2/F_\pi e$. The ratios of masses
and radii $r_B$ for the SK6 and SK4 variants are presented in the last two lines.}\\

The masses of multiskyrmions for the SK6 variant presented in {\bf Table 2} are 
considerably lower than for the SK4 variant (they are between $0.76$ and $0.7$ of 
them, roughly) for accepted choice of parameters. They agree well with
numerical results of recent paper \cite{7}.
For the SK6 variant the $B=1$ configuration can be described by profile of the 
type $(7)$ with 
accuracy about $3\%$, as for the SK4 variant: $M(1)_{appr}\simeq 0.969$ for $b=3.4$
in comparison with numerical value $M(1)=0.940$.
At higher $B$-numbers the upper bound for masses of the SK6 multiskyrmions is not
so close to numerical values as for the SK4 variant. The difference is not smaller
than $2\%$, but from practical point of view it is quite good agreement.
In view of good quantitative agreement of 
analytical and numerical results the studies of basic properties of bubbles of 
matter made in \cite{9} and in present paper are quite reliable.

The width (or thickness) $W$ of the bubble shell can be estimated easily. We 
can define the half-width as a distance between points where $\phi = \pm 1/2$, 
then:
$$ W = 4{r_0 \over b_0} \, ln 3.\eqno (31) $$
Looking at $(6)$ we see that maximal value of $\phi'$ is close to
$b_0/(2r_0)$, and this provides immediately $W \simeq 4r_0/b_0$, in agreement
with $(31)$.
Here there is some difference between the SK4 and SK6 variants of the model,
because the ratio $r_0/b_0$ is different.
For the original SK4 variant at large $B$ the thickness $W \simeq 2\sqrt{2/3}\, ln3$,
i.e. it is universal characteristic of all baryonic bags, and does not depend 
on the number of flavours $N$ as well. For the SK6 variant $W\simeq 
2(8A_N{\cal I}/15B^2)^{1/4}$, i.e. it can depend slightly on $B$ if the ratio
${\cal I}/B^2$ has such dependence, and increases also with increasing number of
flavours. The radius of the bubble grows with increasing $B$ like 
$[{\cal I}/A_N]^{1/4} $ for both SK4 and SK6 variants, see $(15),(16)$.
\section{Toy model: the inclined step approximation}
A natural question is to what extent the "bubble" structure is a necessary
property of multiskyrmions, and what could be instead of this. What we have is
a boundary condition on profile function, $f(0)=\pi$ and $f(\infty) = 0$,
and requirement to get the minimal value of the mass $(3)$. In principle, the
profile $f$ could decrease according to some law different from $(7)$, 
providing another mass and $B$-number distribution, e.g. uniformly filled
bag, and it is just the property of lagrangian $(3)$ that the bubble structure
has an advantage. A good illustration for this provides the toy model of 
"inclined step"  type \cite{4}.
Let $W$ be the width of the step, and $r_0$ - the radius of the 
skyrmion where the profile $f =\pi/2$. $f= \pi/2 - (r-r_o)\pi/W$
for $r_o-W/2 \leq r \leq r_o+W/2 $.
This approximation describes the usual domain wall energy \cite{10}
with accuracy $\sim 9.5$\% .

We can write the energy in terms of $W,\, r_0$, then minimize it
with respect to both of these parameters, and find the minimal value of
energy.
$$ M(W, r_0)={1\over 3\pi}\biggl[{\pi^2 \over W}\biggl(A_Nr_0^2+(1-\lambda )B+
\lambda \frac{3{\cal I}}{8r_0^2}\biggr)+W \biggl(B+(1-\lambda) 
\frac{3 {\cal I}}{8r_0^2}\biggr)\biggr] \eqno (32) $$
This gives
$$W_{min}=\pi\biggl[\frac{A_Nr_0^2+(1-\lambda )B+3\lambda {\cal I}/(8r_0^2)}
{B+(1-\lambda) 3{\cal I}/(8r_0^2) }\biggr]^{1/2} \eqno (33) $$
and the mass
$$M ={2 \over 3}\bigl[\bigl(A_Nr_0^2+(1-\lambda )B+3\lambda {\cal I}/
(8r_0^2)\bigr) \bigl(B+(1-\lambda) 3{\cal I}/(8r_0^2)\bigr)\bigr]^{1/2}.
\eqno (34)$$
The minimization of $(34)$ over $r_0$ should be performed now. For pure SK4
($\lambda =0$) and SK6 variants ($\lambda=1$) it can be made trivially, and
in both cases provides the same result,
 $(r^{min}_0)^2=\sqrt{3 {\cal I} /(8 A_N)}\simeq
0.612 \sqrt{{\cal I}/A_N}$. It is close to the above result $(r^{min}_0)^2(SK4)
 \simeq 0.667 \sqrt{{\cal I}/A_N}$ and $(r^{min}_0)^2(SK6) \simeq 0.73
\sqrt{{\cal I}/A_N}$.
In dimensional units $r^{min}_0= (6{\cal I}/A_N)^{1/4} / (F_\pi e)$.

The thickness of the envelope $W$ is slightly different for both models.
For the SK4 model $W_{min}(SK4)=\pi$ \cite{4}, i.e. it does not depend on $B$ for 
any $SU(N)$, similar to previous result $(31)$ which gives 
$W (SK4)\simeq 1.8 $ for large $B$, 
all in units $2/(F_\pi e)$. For the SK6 variant $W_{min}(SK6) =\pi[3A_N{\cal I}
/2B^2]^{1/4}\, > W_{min}(SK4)$ in this toy model, also similar to result
obtained in previous Section, and increases with 
increasing $N$. Numerically, effect is not great since $A_N < 2$. For $SU(3)$ 
group the factor is $\sim [3A_N/2]^{1/4}=2^{1/4} \simeq 1.19$.

The energy obtained in this way is
$$M_{min}(SK4) \simeq (2B+\sqrt{3A_N {\cal I} /2} ) /3 \eqno (35) $$
and
$$M_{min}(SK6) \simeq {2B\over 3}\biggl(\frac{3A_N {\cal I}}{2B^2}\biggr)^{1/4}
\simeq 0.738\,B \biggl(\frac{A_N{\cal I}}{B^2}\biggr)^{1/4}. \eqno (36) $$

In difference from previous results, $(35)$ and $(36)$ do not give the upper
bound for the skyrmion masses since some terms in expansion in $W$ have been
neglected, and for small $B$, indeed, the value of $(35)$ is
smaller than calculated masses of skyrmions.
For $SU(2)$ model $A_N=1$ and the energy $M_{min}(SK4)= (2B+\sqrt{3{\cal I} /2})/3$.
The formula gives the numbers for $B=3,..., \, 22$ in agreement
with calculation within $RM$ approximation within $2-3 \%$ \cite{3,5}.

More detailed analytical calculation made in present paper confirms the 
results of such 
"toy model" approximation and both reproduce the picture of $RM$ skyrmions as 
a two-phase object, a spherical bubble with profile $f=\pi$ inside and $f=0$ 
outside, and a thickness of the shell which is fixed (the SK4 model), or slightly
depends on $B$-number and $N$ (the SK6 model).

The surface energy density can be estimated. For the SK4 model
$\rho_M^{surf} \simeq (2B+\sqrt{{\cal I}A_N })/ (12\pi r_0^2) \sim 
3\sqrt{A_NB^2/{\cal I}}/8\pi$, or in ordinary units $\rho_M^{surf} \sim 
9\pi\sqrt{A_NB^2/{\cal I}}F_\pi^3e/32$.
The average volume mass density in the shell is, in ordinary units,
$$\rho_M^{vol} \simeq {3\pi \over 64W}(2B+\sqrt{{\cal I}A_N})\sqrt{A_N/{\cal I}}
F_\pi^4 e^2 \sim {9\pi \over 64W}\sqrt{A_NB^2/{\cal I}}F_\pi^4e^2. \eqno (37) $$
For $SU(2)$ model at large $B$ it is about $\sim (0.4\,-\, 0.6)
Gev/Fm^3$ depending on the value of $W$ discussed above, at reasonable choice 
of model parameters $F_\pi=0.186 \,Gev,\; e=4.12$ \cite{4}, i.e. this density 
is several times greater than normal density of nuclei.
For the SK6 model the density of matter in the shell is about $\sim 0.7 - 0.8$ of 
density for the SK4 model.
\section{The properties of the large B multiskyrmions}
It is possible to calculate analytically such characteristics of 
multiskyrmions as mean square radii of baryon number and mass distributions,
tensors of inertia, etc.
Let us estimate first the average dimensions of each cell in the
envelope of the bubble of matter. The average area of the cell is $4\pi r_0^2/
(2B-2)$, at large $B$ it is for the SK4 model
$$S_{cell}\simeq 4\pi \sqrt{{\cal I}/A_N}/(3B).  \eqno (38a) $$
Therefore, the average radius of the cell is
$$ r_{cell} \simeq [{\cal I}/A_N]^{1/4}/\sqrt{3B},\eqno (39) $$
i.e. it depends slightly on the ratio ${\cal I}/B$ which is close to $1$
at large $B$. For the mentioned above choice of $SU(2)$ model parameters
$r_{cell} \sim 0.32 Fm $, about twice smaller than radius of the $B=2$
torus.

For the SK6 variant the dimensions or each cell are slightly greater,
$$S_{cell}\simeq 4\pi \sqrt{2/15}\sqrt{{\cal I}/A_N}/B.  \eqno (38b) $$

The radii of the baryon number and mass distributions can be calculated
analytically for the profile $(7)$ in terms of parameters $r_0,\,b$.
We have
$$ <r_B^2>(r_0,b)\; = \;{2\over \pi} \int r^2s_f^2 f'\,dr \; =\;
{b\over \pi}\int (1-\phi^2)^{3/2} r\, dr \; 
= r_0^2 \frac{b^2-16}{b^2cos (2\pi/b)} \eqno (40) $$
At large $B$,  $<r_B^2>\simeq r_0^2\bigl[1+(2\pi^2-16)/b^2\bigr]$. 
Evidently, when $B\to \infty$ and $b\to \infty $ then
$<r_B^2> \to r_0^2$, as it should be expected.
But for realistic values of $B$ and $b$ the value of $<r_B^2>$ is greater
than $r_0^2$. When $b\to 4, \; (b^2-16)/[b^2cos(2\pi/b)] \to 4/\pi $.

Similarly we can calculate the radii of the mass density distribution.
$$M(B,b)<r_M^2> = {4\over 3} \biggl[T_1 (A_Nb^2/4+2B)+(1-\lambda )T_2 (Bb^2/2+
{\cal I})+\lambda {\cal I}b^2 T_3/4 \biggr]  \eqno (41) $$
$$T_1=\int (1-\phi^2)r^2dr \,=\,\frac{12\pi r_0^3}{b^2sin(3\pi/b)}, $$
$$ T_2=\int (1-\phi^2)^2 dr \,=\, \frac{8\pi
          r_0}{3b^2sin(\pi/b)}(1-1/b^2),
$$
$$T_3 = \int (1-\phi^2)^3/r^2 dr=\frac{32\pi}{15r_0b^2sin(\pi/b)}
\biggl(1-{1\over b^2}\biggr)\biggl(1-{1\over 4b^2}\biggr). \eqno (42)$$
Evidently, at large $B$ $<r^2_M> \to r_0^2$, and $<r_M^2> \simeq <r_B^2>$.

It is possible also to calculate analytically the tensors of inertia of 
multiskyrmion configurations within this approximation. This will be done here
for the SK4 variant which is now of greater practical importance than the SK6 
variant.
The explicit expressions for tensors of inertia of multiskyrmions in general 
$SU(2)$ case are given in \cite{4}. They define the rotation energy of a 
skyrmion in the form
$$ E_{rot} = {1\over 2} \Theta^I_{ab}\omega_a\omega_b +\,
{1\over 2} \Theta^J_{ab}\Omega_a\Omega_b + \,\Theta^{int}_{ab}\omega_a
\Omega_b, \eqno (43) $$
where angular velocities of skyrmions rotations in isotopical $(\omega_a)$ and
usual $(\Omega_b)$ spaces are defined in standard way in terms of corresponding
collective coordinates and their time derivatives, see, e.g., \cite{4} and 
references therein.
The isotopical tensor of inertia
$$\Theta^I_{ab}= \int s_f^2\biggl\{(\delta_{ab}-n_an_b)\biggl[{F_\pi^2\over 4}
+\frac{(\vec{\partial}f)^2}{e^2}\biggr]+{s_f^2\over e^2}\partial_ln_a
\partial_ln_b\biggr\} d^3r  \eqno (44) $$
The expression for the orbital tensor of inertia is much more complicated and
we shall not give it here, see \cite{4} again.
However, the traces of both tensors of inertia are simpler, especially for the
$RM$ ansatz, and depend on the quantities ${\cal N}=B$ and ${\cal I}$ \cite{4}:
$$\Theta^{I}_{aa} = 4\pi \int s_f^2 \biggl\{ {F_\pi^2 \over 2} + {2\over e^2}
\biggl(f'^2+B {s_f^2\over r^2}\biggr)\biggr\} r^2dr ,  $$
$$\Theta^{J}_{aa} = 4\pi \int s_f^2 \biggl\{ B{F_\pi^2 \over 2} + {2\over e^2}
\biggl(Bf'^2+{\cal I} {s_f^2\over r^2}\biggr)\biggr\} r^2dr.\eqno(45)  $$
Evidently, the inequality takes place \cite{4}:
$$\Theta^J_{aa}-B\Theta_{aa}^I ={8\pi \over e^2}({\cal I}-B^2)\int s_f^4 dr
> 0 \eqno (46) $$ 
since ${\cal I} > B^2 $.
The interference tensor of inertia is much smaller for all cases except
the cases of spherical and axial symmetry, as for $B=2$, and will not be
considered here.

The point is that at large $B$ when multiskyrmion is close to spherical bubble
the diagonal components of tensors of inertia can be calculated as $1/3$ of
corresponding traces $(45)$, and off-diagonal tensors of inertia being close
to zero. The accuracy of these statements increases with increasing baryon 
number.

Now we can calculate $(45)$ in our approximation for the profile $(7)$. It
gives:
$$ \Theta^I_{aa}= \frac{4\pi}{F_\pi e^3}\bigl[4T_1+T_2(4B+b^2)\bigr], $$
$$\Theta^J_{aa}= \frac{4\pi}{F_\pi e^3}\bigl[4BT_1+T_2(4{\cal I}+Bb^2)\bigr]
\eqno (47) $$
with $T_1,\,T_2$ given above.
At large baryon numbers $T_1 \simeq 4r_0^3/b$, $T_2 \simeq 8r_0/3b$, and in
natural units for tensors of inertia, $12\pi^2/F_\pi e^3$
$$\Theta^J_{aa}/3 \simeq \frac{16}{27} \sqrt{{2\over 3}} 
 \sqrt{\frac{{\cal I}}{A_N}} \bigl(2B+\sqrt{A_N{\cal I}} \bigr). \eqno(48) $$
So, we obtain 
$$\Theta^J_{aa}/3 \simeq {2\over 3}M_Br_0^2, \eqno (49)$$ 
as it should be for empty spherical bubble with its mass concentrated in
its shell. The equality $(49)$ does not hold, however, for the contributions 
of second order and Skyrme terms in the lagrangian separately. For the 
isotopical tensor of inertia we have inequality
$\Theta^I_{aa}/3 \, < \, 2M_Br_0^2/3B $.
The isoscalar magnetic moment of the baryonic system is
defined by the orbital inertia \cite{4}, $\mu^0 \simeq J B<r_B^2>/3\Theta_J$,
therefore we obtain from $(49)$
$$ {\mu^0\over J} = {B \over 2M_B}, \eqno(50) $$
almost constant for large $B$-numbers.
The consideration of other electromagnetic and weak interaction properties of
multiskyrmions is behind the framework of present paper, it will be made 
elsewhere. 

We finish this Section with a remark that the characteristics
of multiskyrmions obtained here, $<r_B>,\; <r_M>, \; S_{cell}$,
moments of inertia as well as thickness and the mass density of the shell of 
the bubble given in $(31)$ and $(37)$
provide a complete picture of large $B$ multiskyrmions as quasiclassical 
objects formed by the chiral fields.
\section{The role of the mass term} 
Consider also the influence of the chiral symmetry breaking mass term $(M.t.)$
which is described by the lagrangian
$$ -{\cal L}_M =M.t. =\tilde{m} \int r^2(1-cos\,f) dr , \qquad cos\,f=\phi 
\eqno (51) $$
$\tilde{m} = 8\mu^2/(3\pi F_\pi^2 e^2) $, $\mu=m_\pi$. For strangeness, charm, 
or bottom the masses $m_K$, $m_D$ or $m_B$ can be inserted instead of $m_\pi$, 
multiplied by corresponding flavour content of the skyrmion.

Instead of the above expression $(11)$ we obtain now
$$ M_B \, <\, M(B,r_0,b) = \alpha (B,b) r_0 +(1-\lambda)\beta (B,b)/r_0+ 
\lambda \gamma (B,b)/r_0^3 + m \,r_o^3  \eqno (52)$$
with $\alpha,\,\beta,\,\gamma$ given in $(11),(12)$ and 
$m=2\pi\tilde{m}/(b\,sin(3\pi/b))$.
It is possible to obtain in a simple form the precise minimal value of the mass 
for the SK4 model $(\lambda = 0)$
$$M(B,b)={2r_0^{min} \over 3}\bigl(\sqrt{\alpha^2+12m\beta}+2\alpha\bigr)
\eqno(53) $$
at the value of $r_0$
$$r_0^{min}(B,b) =\biggl[\frac{\sqrt{\alpha^2 +12\,m\beta}-\alpha}{6m}
\biggr]^{1/2}. \eqno (54) $$
Eq-ns. $(52)$ and $(53)$ give the upper bounds for the mass of the
skyrmion,
because they are calculated for the profile $(7)$ different from the true
profile which can be obtained by explicit minimization of the energy 
functional $(3)$ with the mass term included. In particular, it is well known
that the tail of the profile decreases exponentially, $ \sim exp(-\mu r)/r$ at
$r \,>\,\sqrt{2B}/\mu $ instead of the power law. However, for multiskyrmions 
at large $B$ the main contribution to the mass, moments of inertia of the skyrmion,
etc. is due to the shell of the bubble, and
the relative contribution of the tail (i.e. the region outside of the
bubble) decreases as $1/\sqrt{B}$, at least, being not important at large $B$. 

When the mass $m$ is small enough, as for the pion, the expansion in 
$12m\beta/\alpha^2$ can be
made, and one obtains the following reduction of the skyrmion size $r_0$:
$$ r_0 \to r_0 - {3m\over 2\alpha} \biggl(\frac{\beta}{\alpha}\biggr)^{3/2}
\simeq \sqrt{{2\over 3}}\biggl(\frac{{\cal I}}{A_N}\biggr)^{1/4}\biggl[1-\,
\frac{3\pi m}{2(2B+\sqrt{{\cal I}A_N})}\biggl(\frac{{\cal I}}{A_N}\biggr)
^{3/4}\biggr], \eqno (55) $$
and increase of the soliton mass
$$\delta M = M \frac{m\beta}{2\alpha^2}\biggl[1-\frac{9m\beta}{8\alpha^2}
\biggr]\simeq M \,m\frac{\pi ({\cal I}A_N)^{3/4}}{2(2B+\sqrt{{\cal I}A_N}).}
 \eqno (56) $$
We used that at large $B$
$$\alpha \simeq {1\over 3\pi}\biggl(A_Nb+{8B\over b}\biggr) \sim \sqrt{B}, 
\qquad \beta \simeq {4\over 9\pi}\biggl(Bb + \frac{2{\cal I}}{b}\biggr) \sim
B^{3/2}, \qquad 
\gamma ={8 \over 45\pi} {\cal I} \sim B^2. \eqno (57) $$
As it was expected from general grounds, dimensions of the soliton decrease
with increasing $m$.
However, even for large value of $m$ the structure of multiskyrmion at 
large $B$
remains the same: the chiral symmetry broken phase inside the spherical
wall where the main contribution to the mass and topological charge is 
concentrated \cite{4}. The value of the mass density inside of
the bubble is defined completely by the mass term with $1-\phi= 2$.
The baryon number density distribution is quite similar, with only difference
that inside the bag it equals to zero. Eq. $(52),(53)$ provide the upper bound
for the skyrmion mass, the lower bound $(4a)$ takes place again for the mass
without the mass term.
It follows from these results that $RM$ approximated multiskyrmions cannot
model real nuclei at large $B$, probably $B > 12 -20$, and configurations like
skyrmion crystals \cite{11} may be more valid for this purpose.

One of the issues of multiskyrmions phenomenology are the multibaryons with 
flavour different from that of $u,d$ quarks, in particular $s,c$ and $b$ flavours,
see \cite{12,4} and references therein. The important ingredient of the calculation of spectra of such multibaryons is
calculation of the flavour excitation energies $\omega_{B,s},\, \omega_{B,c}$
and $\omega_{B,b}$. The behaviour of these energies as a function of the 
baryon number
is important for the conclusion if the corresponding flavour is more bound when
$B$ increases, or less bound.
The following expression for these energies was obtained (\cite{12,4} and 
references therein):
$$\omega_{B,F} =  \frac{N_cB}{8\Theta_{F,B}}(\mu_{F,B}-1) \eqno (58) $$
with
$$\mu_{F,B}\simeq [1+16\bar{m}_D^2\Gamma_B\Theta_{F,B}/(N_cB)^2]^{1/2} 
\eqno (59)$$
and the $\sigma$-term $\Gamma_B$ and "flavour" moment of inertia $\Theta_{F,B}$
given by
$$\Gamma_B ={F_\pi^2\over 2}\int (1-c_f) d^3r,\qquad
\Theta_{F,B} = {1\over 8}\int(1-c_f)\biggl[F_D^2+{1\over e^2}\bigl(
(\vec{\partial}f)^2+s_f^2(\vec{\partial}n_i)^2\bigr)\biggr], \eqno (60)$$
$N_c$ is the number of colours of the underlying QCD, the last term in $(58)$,
proportional to $N_c$, is due to Wess-Zumino term present in the action of the
model.
We neglected the terms of relative order $(F_D^2-F_\pi^2)/m_D^2$ because,
as it was shown explicitly in \cite{12}, they are important only for small 
values of $B$, and make vanishing contribution for the large $B$ which we 
consider here.
The terms $\sim (1-c_f)$ in the integrand of $(60)$ give contribution 
proportional to the volume of skyrmion, $\sim r_0^3$, whereas terms $\sim s_f^2$ or $(\partial
F)^2$, etc, are due to the shell of the skyrmion, only, $\sim r_0^2$.
$\bar{m}_D^2=F_D^2m_D^2/F_\pi^2-m_\pi^2$, $m_D$ is the mass of flavoured
meson ($K,\, D$ or $B-$ meson). When $\bar{m}_D$ is sufficiently large - 
practically it is always fulfilled - then
$$\omega_{F,B} \simeq {\bar{m}_D\over 2} \biggl(\frac{\Gamma_B}{\Theta_{F,B}}
\biggr)^{1/2} -\frac{N_cB}{8\Theta_{F,B}}\eqno (61) $$
For small $B$
the energies decrease somewhat with increasing $B$, and this leads to the
increase of binding energies of flavours. From the picture of
large $B$ multiskyrmions clarified here we can see that the quantity $\Gamma_B$
has the part proportional to the volume occupied by multiskyrmion, i.e.
$\sim B^{3/2}$, whereas for $\Theta_{F,B}$ this part is considerably smaller. 
As a result, at large $B$ the ratio
$$\frac{\Gamma_B}{\Theta_{B,F}} \to 4 {F_\pi^2 \over F_D^2} \,=\,
\frac{4}{\rho_D^2}\eqno (62) $$
For the difference $\epsilon_{F,B}=m_D-\omega_{F,B}$ which is important
contribution into the binding energy of flavoured meson by $SU(2)$ skyrmion
we obtain
$$\epsilon_{F,B} \simeq {1\over 2}\frac{m_\pi^2}{m_D \rho_D^2} + 
2m_D\frac{\Theta^{surf}_{F,B}}{\rho_D^2 \Gamma_B}+
\frac{N_cB}{2\Gamma_B\rho_D^2},\eqno (63) $$
where the surface, or shell, contribution to flavour inertia 
$\Theta^{surf}_{F,B}$ is proportional to $1/e^2$ in $(60)$.
Only the first term in $(63)$ comes from the volume of the multiskyrmion, but
numerically it is negligible, except the case of the strangeness. Remaining
terms are both of the surface (or shell) origin,
in other words, the binding of heavy flavour takes place on the shell of the
multiskyrmion. The second term, $\sim \Theta^{surf}_{F,B}/\Gamma_B$, which
dominates in magnitude, decreases with
increasing $B$ as $1/\sqrt{B}$, and this explains the results of calculations
\cite{4,12} which have shown that the binding of flavour becomes weaker with
increasing $B$. However, this property may be intrinsic for $RM$ 
multiskyrmions, and can be absent for the skyrmion crystals, for example.
\section{Discussion and conclusions}
In \cite{9} and in the present paper we established the 
link between the topological soliton models in rational maps approximation, for two
different modifications of the Skyrme model, the SK4 and SK6, and
the soliton models of "domain wall" type.
It was possible by means of consideration of a class of simple functions $(7)$
approximating both the skyrmion profiles and the spherical domain walls. The 
ansatz $(7)$ does not describe correctly the asymptotical behaviour of the 
profile at large distances, although provides quite accurate description of 
many characteristics of multiskyrmions.
This simple picture of the $RM$ multiskyrmions allows to understand
some peculiarities of multiskyrmions phenomenology which appeared as a result 
of the calculations, or as computer evidence.

The upper bound for the energy of multiskyrmions is obtained which is very
close to the known energies of the $RM$ multiskyrmions, especially at largest $B$,
and is higher than the known lower bound by $\sim 4\%$ only for the SK4 (pure
Skyrme) variant of the model, and by $\sim 9\%$ for the SK6 variant. For the 
SK6 variant the upper bound obtained is
considerably lower than for the SK4 variant, dimensions of solitons are larger by 
few $\%$, at asymptotically large baryon numbers for accepted choice
of model parameters.

The following properties of bubbles of matter from $RM$ multiskyrmions are
established analytically, mostly independent on particular values of the
quantity ${\cal I}$:

The dimensions of the bubble grow with $B$ as $\sqrt{B}$, or as 
${\cal I}^{1/4}$, whereas the mass is proportional to $\sim B$ at large $B$.
Dimensions of the bubble decrease slightly with increasing $N$ - the
number of flavours, $r_0 \sim [N/\bigl(2(N-1)\bigr)]^{1/4}$,
see $(15),(16)$.

For the SK4 variant the thickness of the bubbles envelope $(31)$ is constant 
at large $B$ and does not depend on the number of flavours, therefore, the 
average surface mass density is constant at large $B$, as well as average 
volume density of the shell $(37)$. Both densities increase slightly with 
increasing 
$N$. For the SK6 variant the thickness of the shell slightly depends on the
ratio ${\cal I}/B^2$. At the same time the mass and $B$-number
densities of the whole bubble $ \to 0$ when $B\to \infty$, and this is in 
contradiction with nuclear physics data confirming the constant density of
nuclear matter. The bubble structure of multiskyrmions is more pronounced for
the SK4 variant of the model.
The material which the shell of the bag is made of looks like honeycomb,
or web with constant average area of each cell.

The treatment performed in the present paper could be generalized in several 
directions. More general effective lagrangians can be considered, including
higher terms in chiral derivatives, and the bundle of profiles can be
introduced instead of one, see $(7)$:
$$ \phi = \int \rho (b) \frac{(r/r_0)^b-1}{(r/r_0)^b+1} db \eqno (64) $$
with the evident normalization condition $\int \rho (b) db = 1$.
The multiskyrmion mass can be written then in the form:
$$M(B) = \mu_2(B) r_0 + \mu_4(B)/r_0 + \mu_6(B)/r_0^3 +\mu_8(B)/r_0^5 +...
\eqno (65) $$
with $\mu_2(B) = \int \rho(b) \alpha (B,b) db$, etc. It would be of interest 
to study the advantages and prospects of such more general approach.
Some models for higher order terms in the lagrangian have been considered
in \cite{13} for the $SU(2)$ skyrmions.

It follows from the above consideration that the spherical bubble or bag 
configuration can be obtained from
the lagrangian written in terms of chiral degrees of freedom only, i.e.
the Skyrme model lagrangian leads at large baryon numbers to the formation 
of spherical bubbles of matter and thus provides a field-theoretical
realization of the bag-type model. In such models, which have been popular
a time ago, the properties of the building material of bags have been
postulated \cite{14}. It was the technical obstacle in formulating the bag 
model as a
consistent quantum theory: the bag boundary $R(t)$ was prescribed externally,
and special efforts have been made to overcome this \cite{15}, not successful
completely. The chiral soliton models leading to existence of multiskyrmions
are free of this drawback.
It should be noted that there is difference of principle between hadronic bags
of the MIT type and the bubbles of matter which appear in effective field 
theories as $RM$ multiskyrmions. In the first case the bag is the region where 
the quark-gluon phase of matter is restricted, in the latter case it is the region
with skyrmion profile function $f=\pi$, different from its vacuum value
$f=0$. The quark-gluon phase does not enter the
consideration at all, although it can exist inside the skyrmionic bubble,
due to the known Cheshire Cat principle \cite{16,17}.

This picture of the mass and $B$-number distribution in the $RM$ multiskyrmions 
condradicts to what is known about nuclei,
however, it emphazises the role of periphery of the nucleus and could be an
argument in favour of shell-type models of nuclei. The skyrmion crystals 
\cite{11} are believed to be more adequate for modelling nuclear matter.

It would be of interest to perform the investigation of the dynamics of
bubbles in the chiral soliton models similar to that performed recently for
the simplified two-component sigma model, or the sine-Gordon model in $(3+1)$ 
dimensions \cite{18}.
Observations concerning the structure of large $B$ multiskyrmions made here
can be useful in view of possible cosmological applications of Skyrme-type
models, see e.g. \cite{19}. The large scale structure of
the mass distribution in the Universe \cite{20} is similar to that in 
topological soliton models, and it can be the consequence of the similarity of 
the laws in micro- and macroworld.
We conclude with a remark that analytical methods which are not typical for
studies of skyrmion properties (see also \cite{21}) allow to obtain
very simple and transparent results which accuracy increases with increasing
baryon number.

As it was noted by referee, the parametrization of the profile function
similar to $(7)$ was proposed long ago in \cite{22} for $B=1$ hedgehog.
It was $ f(r,r_0) = 2\,atan (r_0/r)^2$, or 
$$cos\,f= \frac{(r/r_0)^4-1}{(r/r_0)^4+1},$$ 
which has correct asymptotics at large $r$ with the appropriate choice of $r_0$, 
but does not minimize the mass of the skyrmion. Parametrization 
$(7)$ with the power $b$ as a parameter to be fitted provides description
of masses especially good for the large baryon numbers.

I would like to thank W.J.Zakrzewski, T.Ioannidou for interest in the 
questions discussed in present paper and useful comments. I'm indebted to 
P.Sutcliffe and B.Piette for sending me the results of their
investigations.
This work is supported by the Russian Foundation for Basic Researches,
grant RFBR 01-02-16615. \\
\vglue 0.2cm
{\elevenbf\noindent References}
\vglue 0.1cm

\end{document}